\documentclass[letter]{aa}  
\usepackage{graphicx}
\usepackage{epstopdf}
\usepackage{txfonts}
\begin{document}

   \title{Accretion from debris disks onto white dwarfs}

   \subtitle{Fingering (thermohaline) instability and derived accretion rates}

   \author{M. Deal \inst{1, 2},
          S. Deheuvels\inst{1, 2}, G. Vauclair \inst{1,2}, S. Vauclair \inst{1, 2}, \and F. C. Wachlin \inst{3, 4}
          }
         \institute{
           Universit\'e de Toulouse, UPS-OMP, IRAP, France,         
         \and
             CNRS, IRAP, 14 avenue Edouard Belin, F-31400 Toulouse, France,
         \and
             Facultad de Ciencias Astronomicas y Geofisicas, Argentina (FCAG, UNLP),
         \and
             Instituto de Astrofisica de La Plata, Argentina (IALP, UNLP-CONICET),
             }

   \date{Received 4 July 2013 / accepted 11 August 2013 }

 
  \abstract
   {Recent observations of a large number of DA and DB white dwarfs show evidence of debris disks, which are the remnants of old planetary systems. The infrared excess detected with \emph{Spitzer\/} and the lines of heavy elements observed in their atmospheres with high-resolution spectroscopy converge on the idea that planetary material accretes onto these stars. Accretion rates have been derived by several authors with the assumption of a steady state between accretion and gravitational settling. The results are unrealistically different for DA and DB white dwarfs. }
   {When heavy matter is accreted onto stars, it induces an inverse $\mu$-gradient that leads to fingering (thermohaline) convection. The aim of this letter is to study the impact of this specific process on the derived accretion rates in white dwarfs and on the difference between DA and DB.}
   {We solve the diffusion equation for the accreted heavy elements with a time-dependent method. The models we use have been obtained both with the IRAP code, which computes static models, and the La Plata code, which computes evolutionary sequences. Computations with pure gravitational settling are compared with computations that include fingering convection.   }
   {The most important result is that fingering convection has very important effects on DAs but is inefficient in DBs. When only gravitational settling is taken into account, the time-dependent computations lead to a steady state, as postulated by previous authors. When fingering convection is added, this steady state occurs much later.}
   {The surprising difference found in the past for the accretion rates derived for DA and DB white dwarfs disappears. The derived accretion rates for DAs are increased when fingering convection is taken into account, whereas those for DBs are not modified. More precise and developed results will be given in a forthcoming paper. }


   \maketitle
%

\section{Introduction}
Recent high-resolution spectroscopy observations have revealed the presence of heavy elements in the spectra of an increasing number of DA and DB white dwarf stars,
classified as DAZ and DBZ white dwarfs. They account for a fraction  
between 5 (\citealt{koester05}) and 20 percent (\citealt{zuckerman03}) of the DA and DB white dwarfs with effective temperatures below 25000K (\citealt{zuckerman07}, 
\citealt{klein10a}, 
\citealt{venne10}, \citealt{zuckerman11}, \citealt{dufour12}). When different chemical species are present in the stellar atmosphere, their abundance ratios
are similar to the abundance ratios determined for the Earth's bulk (\citealt{melis11}). In the meantime, observations from \emph{Spitzer\/} revealed an infrared excess 
from a number of white dwarfs. This excess is interpreted as the proof of a circumstellar disk orbiting these stars (\citealt{farihi11}, \citealt{xu12}, 
\citealt{girven12}, \citealt{brinkworth12}). Many of the white dwarfs exhibiting such infrared excess also show heavy elements in their atmosphere. These two observational facts are linked and indicate that material from the disk is accreted onto the white dwarfs, polluting their atmospheric composition with heavy elements. These disks seem to be the remnant of  
planetary systems orbiting the white dwarfs. 

   The presence of heavy elements in white dwarf atmospheres raises the question of how they can remain at the surface, either in the radiative photophere or in the surface 
convection zone since the gravitational settling takes place on a very short time scale.
As shown by \cite{vauclair79} and \cite{chayer95}, the radiative levitation is not efficient enough to balance this settling for effective temperatures below 25000~K. This suggests that we observe an ongoing accretion process of disk material, which may be due to the tidal disruption of an asteroid-like body (\citealt {jura03}) responsible for the infrared excess.
  
   The accreted heavy elements diffuse downwards either at the radiative photospheric level or below the convective zone inside which they have been  mixed. A number of
studies have estimated the accretion rate by using the observed heavy element abundances, assuming a steady state between the accretion rate and the 
gravitational settling. Up to now, these estimates have only taken the gravitational settling into account below the standard convective zones (\citealt{dupuis92I}, 
\citealt{dupuis93II}, \citealt{dupuis93III}, \citealt{koester09}, \citealt{farihi12}). Using this method, \cite{farihi12} obtained accretion rates ranging from $10^{5.5}$ to $10^{9.5}$ g.s $ ^{-1}$ for DAZ and from $10^{5.5}$ to $10^{11.5}$ g.s$^{-1}$ for DBZ white dwarfs. These values correspond to a few asteroid masses accreted in millions of years. More precisely, $10^{9.5}$ g.s $^{-1}$ accreted during ten million years are equivalent to the mass of Ceres. That accretion rates above $>10^{9.5}$ g.s $^{-1}$ 
are only found for DBZ is surprising since the white dwarf envelope composition should not play any role in the physics of the accretion process.
 
\cite{deal13} points out that an efficient physical mechanism had been forgotten in these computations, namely "fingering" convection, which is similar to the thermohaline convection observed in the oceans (see, for example, \citealt{huppert76} and \citealt{gill82}). This instability is likely to occur whenever heavy elements accumulate in the outer layers of a star (\citealt{vauclair04}, \citealt{garaud11},\citealt{theado12bis})
   
In this letter we explore the consequences of this process on the derived accretion rates. We show the results obtained by solving 
the diffusion equation for heavy elements, including both the gravitational settling and the fingering convection, for a few typical models of DA and DB white dwarf stars. We find that the fingering convection plays a dominant role for the DAZ but has a negligible 
impact on the DBZ. This process, forgotten in previous computations, may reconcile the accretion rates derived for DAZ and DBZ white dwarfs. 


\section{Global treatment of fingering convection and gravitational settling}

\subsection{The fingering (thermohaline) convection}

Accumulating heavy elements in the outer layers of a star induces an inverse $\mu$ gradient in the radiative zone below the mixed region. This accumulation leads to fingering (thermohaline) instability, provided that the so-called "density ratio", defined as  $R_{0} = \frac{\nabla_{ad}-\nabla}{|\nabla_{\mu}|}$ , is larger than one and smaller than the Lewis number, ratio of the thermal to the atomic diffusivities. Here the gradients have their usual meanings: the adiabatic and real thermal 
gradients ($\nabla=\frac{\partial ln T}{\partial lnP}$, $\nabla_{ad}=\left(\frac{\partial ln T}{\partial lnP}\right)_{ad}$)
 and the $\mu$-gradient ($\nabla_\mu=\frac{\partial ln \mu}{\partial lnP}$).  

Fingering convection is usually treated as a macroscopic diffusion process, with a mixing coefficient that has recently been derived in various ways. Here we use the expression given by \cite{vauclair12}, calibrated on the 3D numerical computations by \cite{TGS11}:
 \[
  D_{th}=C_{t}\kappa_{t}(R_{0}-1)^{-1}(1-R_{0}\tau)
   \]
where $C_t=12$ (calibrated constant), $\kappa_t$ is the thermal diffusivity, and $\tau$ the inverse Lewis number.
  
\subsection{The diffusion equation}

The general equation of diffusion, including gravitational (microscopic) settling and fingering (macroscopic) mixing is written as
    \[
\frac{\partial (\rho c)}{\partial t}+ div(\rho c (V_D + V_{th}))=0 
    \]
   where $\rho$ is the density, $V_D$ the microscopic diffusion velocity, $V_{th}$ the fingering diffusion velocity, and $c$ the number concentration of heavy elements.

The microscopic diffusion velocity is computed in the same way as usual (see, for example,\citealt{theado12bis}), and the fingering velocity is given by
   \[
   V_{th} = -D_{th}\frac{1}{c}\frac{\partial c}{\partial r},
   \]
   where $D_{th}$ is the mixing diffusion coefficient.

\section{Models}

\subsection{Choice of the DA and DB models}

We solved the diffusion equation either with gravitational settling alone or with both gravitational settling and fingering convection in a sample of DA and DB white dwarf models.  We selected six models with parameters representative of six observed white dwarfs covering the range of effective temperatures of the DAZ and DBZ (see Table \ref{models}). We used two sets of DA and DB models with similar effective temperatures of 10600~K and 17000~K, respectively, to compare the effect of fingering convection for stars with different chemical compositions and the same temperature, and for stars with same chemical composition but different temperatures. We also added one DA model with an effective temperature of 12800~K, which is representative of white dwarfs with thin convective zones, and one DB with an effective temperature of 21110~K as an example of hot DBs. The mass of the models has been chosen as 0.59~$M_{\odot}$, which is typical of the observed white dwarfs.

\subsection{Characteristics of the models}

The DA models were computed with the IRAP code originally described in \cite{dolez81} and updated by more recent opacity tables, equation of state, and electron conduction 
from \cite{itoh92}. For this preliminary study we only considered DA models with thick hydrogen mass fraction of $M_{H}/M_{*}$= $10^{-5}$. For the DB, we selected models in an evolutionary sequence obtained with the stellar evolution code of La Plata (LPCODE; \citealt{althaus10b}). 
 
 \begin{table}
   \caption{Models of DA and DB white dwarfs.}             
   \label{models}      
   \centering                   
   \begin{tabular}{c r r r r }    
   \hline  
   \noalign{\smallskip}               
   Models & $T_{eff}$ [K] & Log g & $M_{cz}$ [g] & Type \\    
   \hline
   \noalign{\smallskip}                        
      1 & 10600K & 7.90 & 1.68 $10^{21}$ & DAZ \\      
      2 & 12800K & 7.88 & 1.82 $10^{17}$ & DAZ \\
      3 & 16900K & 7.80 & *3.52 $10^{16}$ & DAZ \\
      4 & 10600K & 8.02 & 1.84 $10^{28}$ & DBZ \\
      5 & 17100K & 7.99 & 1.39 $10^{26}$ & DBZ \\
      6 & 21110K & 7.98 & 2.70 $10^{24}$ & DBZ \\
   \hline                                   
   \end{tabular} 
   \tablefoot{$M_{cz}$ is the mass contained in the convective zone. (*)This mass is arbitrarily defined because the model has no convective zone.}  
   \end{table}

\subsection{The computations}

The diffusion equation was solved using a semi-implicit finite differences method as described by \cite{charbonnel92}. Because of the very different characteristic time steps for the two processes, and the non-linearity of the mixing by fingering convection, the computations were quite time-consuming. For this reason, we used static models without taking the increase in the depth of the dynamical convective zone induced by the accumulation of heavy elements into account. One of us (F.W.) began a full treatment of this mixing along a white dwarf evolution sequence. Preliminary results show that the fractional mass mixed by the fingering convection exceeds the increase in the dynamical convective mass by orders of magnitude. Results obtained with this full treatment will be included in a forthcoming paper. 

\subsection{The treatment of accretion}

We considered a continuous accretion rate of an average fictitious element with a mass of 40 corresponding to calcium. We converted the observed Ca
abundance into total accreted mass by assuming that Ca represents 1.6 percent of the accreted mass in agreement with the Earth's bulk composition (\citealt{allegre95}), as in \cite{farihi12}. The calcium values that have been chosen as constraints for the models are consistent with the abundances observed for stars of similar effective temperatures by \cite{koester06} for the DAZ, and \cite{zuckerman10} and \cite{desharnais08} for the DBZ white dwarfs.

The accretion was assumed to occur isotropically on the white dwarf surface. When there is a surface convective zone, we considered that the accreted mass is mixed in 
the convective zone instantaneously. This is justified since the typical convective turnover time scale is a few seconds in white dwarfs. When there is no surface convection zone, we assumed that the accreted material is mixed in an arbitrary surface layer corresponding to the photosphere, as in \cite{koester09}. 
 
\section{The results}

\subsection{Determination of accretion rates}

To determine the accretion rates, we iterated our computations each time so as to obtain the same element abundances in the outer layers of the models as observed. 
Our results show that in most cases, as postulated by \cite{farihi12}, a steady state may actually be reached after some time; however, this happens much later when fingering convection is added rather than for gravitational diffusion alone.

An interesting case is that of DA white dwarfs with very thin convective zones. This case is discussed in the conclusions below, and presented in more detail in a forthcoming paper.

\subsection{Results for the DAs}

\begin{figure}
   \centering
   \includegraphics[width=8cm]{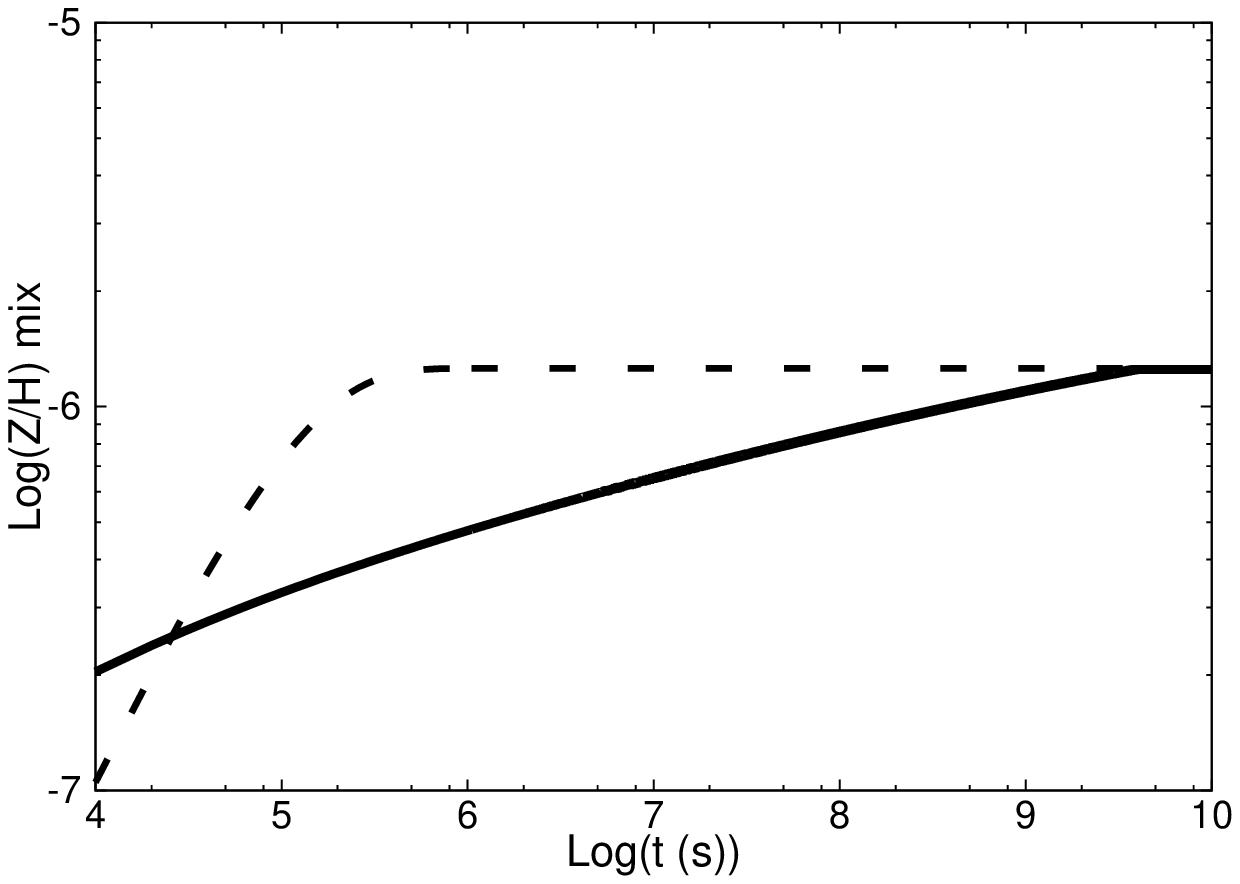}
   \includegraphics[width=8cm]{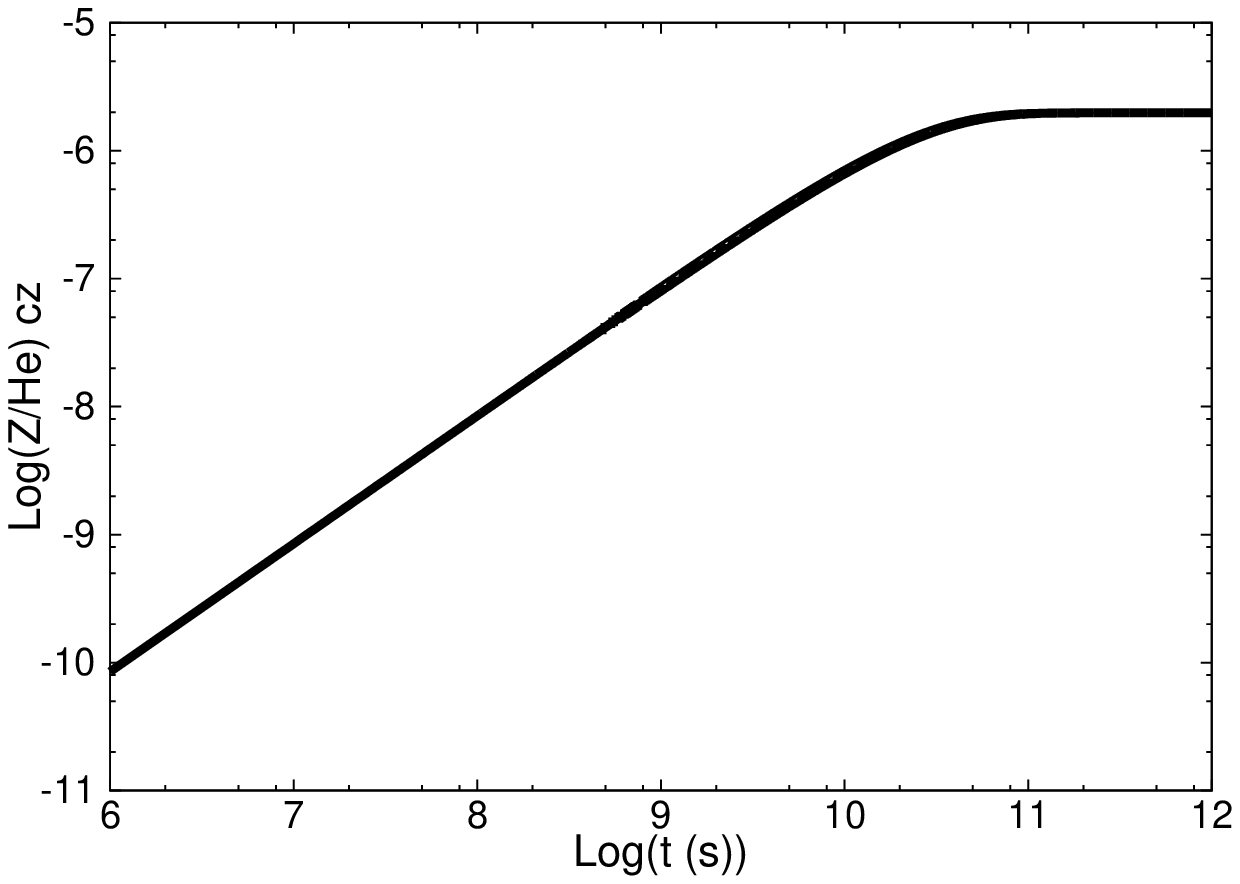}
   \caption{Time evolution of the surface metal abundance by number of hydrogen (or helium) for the DA model 3 (top panel) and the DB model 5 (bottom panel) for accretion rates listed Table \ref{accr}. The dashed lines represent the behaviour with only gravitational settling taken into account, and the solid lines represent the behaviour with both fingering convection and gravitational settling. The two curves are superposed for the DB model where fingering convection is inefficient.} 
   \label{M/H(t)}
   \end{figure}

\begin{figure}
   \centering
   \includegraphics[width=8cm]{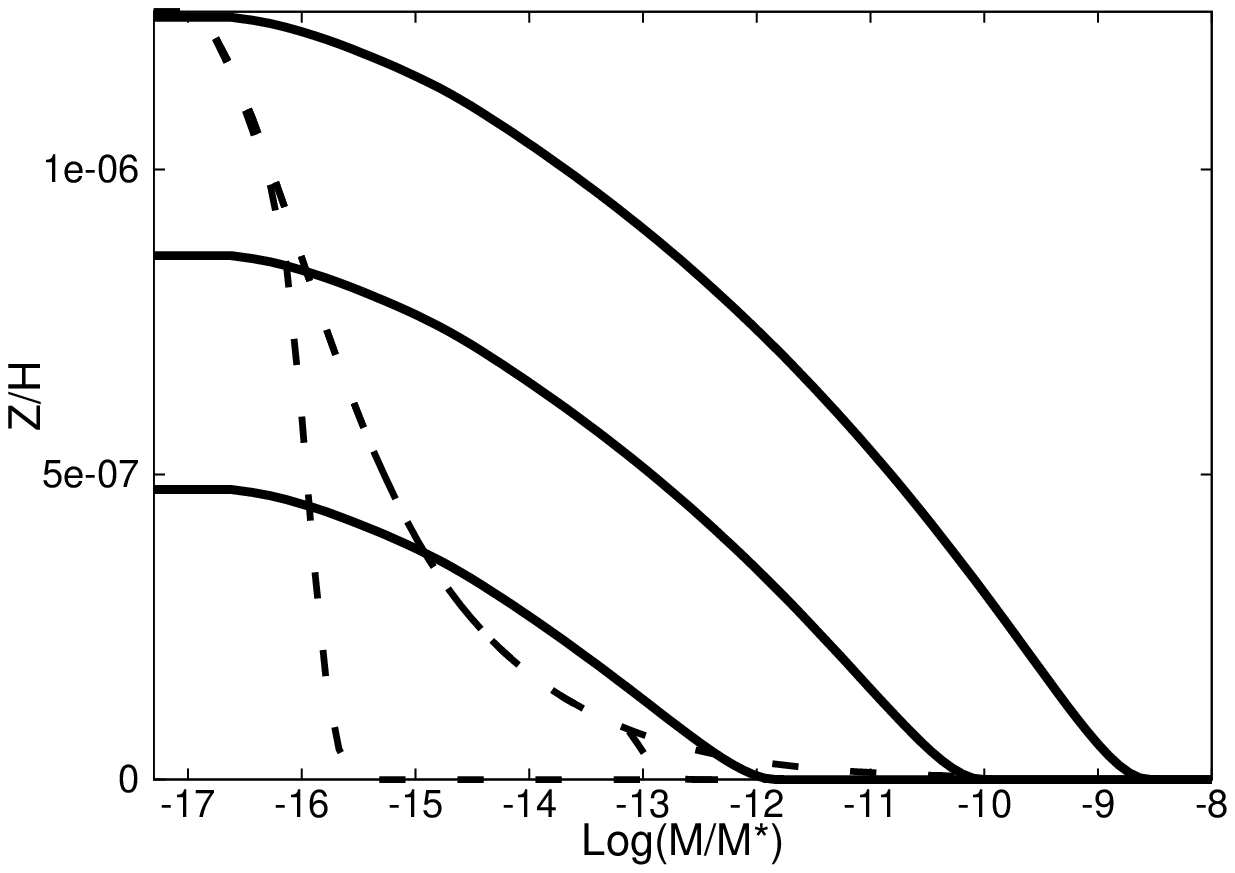}
   \includegraphics[width=8cm]{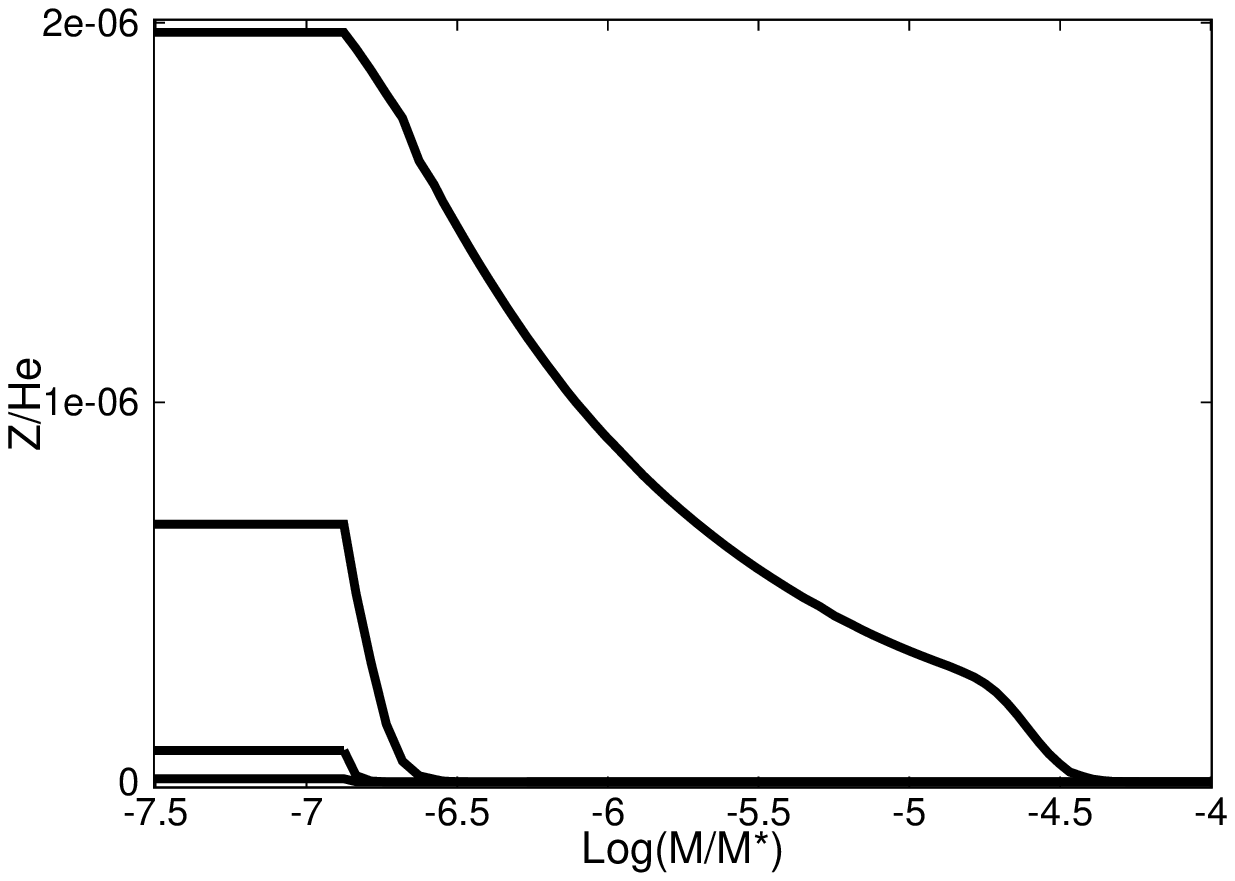}
   \caption{Time evolution of the surface metal abundance profiles by number of hydrogen (or helium) according to the depth in the star for the DA model 3 (top panel) and the DB model 5 (bottom panel) for accretion rates listed Table \ref{accr}. From bottom to top there are chronological times (t=$10^6$s, t=$10^8$s, and t=5 $10^9$s for the DA model, and t=$10^8$s, t=$10^9$s, t=$10^{10}$s, and t=$10^{12}$s for the DB model). The dashed lines represent the behaviour with only gravitational settling taken into account, and the solid lines represent the behaviour with both fingering convection and gravitational settling. For the DB model profiles are identical because fingering convection does not develop.}
   \label{profiles}
   \end{figure}

To compare our work with the previous ones, we performed two types of computations, first with gravitational settling alone, and second with both gravitational settling and fingering convection. The results are presented in the top panel of Fig. \ref{M/H(t)} for the DA model 3 with 16900~K.  When fingering convection is included, the accretion rates needed to reproduce the observed surface abundances are larger by almost 2 orders of magnitude (see Table \ref{accr}). A steady state is reached in both cases but at a later time when fingering convection is included. The abundances profile shapes evolve very differently as shown in Fig. \ref{profiles} (top panel). The fingering convection mixes the heavy elements much more deeply than predicted when only gravitational settling is taken into account.  

\subsection{Results for the DBs}

The same computations have been done for the DB model 5 with 17100~K (bottom panel Fig. \ref{M/H(t)}). In this case, the fingering convection is inefficient, and the two curves showing the time evolution with and without the fingering convection are indistinguishable. The accretion rates are the same (see Table \ref{accr}). The  profile shapes are also similar (Fig. \ref{profiles} bottom panel).  The inefficiency of the fingering convection in the DBs is due to a $R_0$ value higher than the Lewis number below the convection zone, which precludes the fingering instability developing. Only for the  hottest  DB model (model 6 with 21110~K) does the fingering convection begin to play a marginal role. 

   \begin{table}
   \caption{Accretion rates (g.s$^{-1}$) and observed heavy element abundances by number of hydrogen (or helium). (a) Accretion rates computed with the assumption of pure gravitational settling; (b)Accretion rates computed with fingering convection included.}             
   \label{accr}      
   \centering                   
   \begin{tabular}{c r r r r r}    
   \hline  
   \noalign{\smallskip}               
   Models & Log$\dot{M} (a)$ & Log$\dot{M} (b)$ & Log[Ca/H(e)] & Log[Z/H(e)]\\    
   \hline
   \noalign{\smallskip}                              
      1 (DAZ) & 9.23 & 9.83 & -7.2 & -5.4 \\
      2 (DAZ) & 7.74 & 10.01 & -7.1 & -5.3 \\
      3 (DAZ) & 7.70 & 9.40 & -7.7 & -5.9 \\
      4 (DBZ) & 8.04 & 8.04 & -12.4 & -10.6\\
      5 (DBZ) & 10.08 & 10.08 & -7.5 & -5.7 \\
      6 (DBZ) & 10.70 & 10.80 & -8.0 & -6.2 \\
   \hline                                   
   \end{tabular}
   \end{table}
   
\section{Discussion and conclusion}

Our most important result is the strong behaviour difference of fingering convection in DA and DB white dwarfs. For similar effective temperatures, it plays a major role in the hydrogen environment of DAs but is completely inefficient in the helium environment of DBs. The reasons include the difference in the depth of convective zones, the difference in the thermal diffusivity in the radiative zone below, and the difference in the ambient mean molecular weight. Although more computations are needed to study all the white dwarfs with observed heavy elements, we strongly suggest that the difference in the derived accretion rates for both kind of stars, as found by previous authors, comes from not taking fingering convection into account.

Our numerical computations confirm that a steady state between accretion and diffusion  may generally be reached after some time, and much later when fingering convection is considered than with pure gravitational settling. One interesting case, not shown in this paper, but one that will be extensively presented in a forthcoming paper, is that of DAs with a very small hydrogen layer. In this case the accreted matter is mixed by fingering convection down to the transition region between H and He. Owing to the stabilizing $\mu$-gradient, fingering convection is stopped at that depth. The only reason that heavy elements may still migrate below this region is their microscopic settling in the helium environment, which is quite slow. For this reason, and only in this case, heavy matter remains in the stellar outer layers long after accretion has stopped.

In all other cases, accretion rates may be derived in the same way as previously done with pure settling, except that the results are several orders of magnitude more than before for DAs. Fingering convection seems to be the missing process in previous computations, which reconciles the rates derived for DAs and DBs. This conclusion is quite satisfying because no reasons could be found for a different accretion process according to the internal chemical composition of the accreting star. We now know that the same accretion process can account for the observations of both kinds of stars.

\begin{acknowledgements}

We thank Noel Dolez for providing models of DA white dwarfs, computed with the IRAP code.
     
\end{acknowledgements}

\bibliographystyle{aa.bst} 
\bibliography{biblio} 

\begin{thebibliography}{36}
\expandafter\ifx\csname natexlab\endcsname\relax\def\natexlab#1{#1}\fi

\bibitem[{{All{\`e}gre} {et~al.}(1995){All{\`e}gre}, {Poirier}, {Humler}, \&
  {Hofmann}}]{allegre95}
{All{\`e}gre}, C.~J., {Poirier}, J.-P., {Humler}, E., \& {Hofmann}, A.~W. 1995,
  Earth and Planetary Science Letters, 134, 515

\bibitem[{{Althaus} {et~al.}(2010){Althaus}, {C{\'o}rsico}, {Bischoff-Kim},
  {Romero}, {Renedo}, {Garc{\'{\i}}a-Berro}, \& {Miller
  Bertolami}}]{althaus10b}
{Althaus}, L.~G., {C{\'o}rsico}, A.~H., {Bischoff-Kim}, A., {et~al.} 2010,
  \apj, 717, 897

\bibitem[{{Brinkworth} {et~al.}(2012){Brinkworth}, {G{\"a}nsicke}, {Girven},
  {Hoard}, {Marsh}, {Parsons}, \& {Koester}}]{brinkworth12}
{Brinkworth}, C.~S., {G{\"a}nsicke}, B.~T., {Girven}, J.~M., {et~al.} 2012,
  \apj, 750, 86

\bibitem[{{Charbonnel} {et~al.}(1992){Charbonnel}, {Vauclair}, \&
  {Zahn}}]{charbonnel92}
{Charbonnel}, C., {Vauclair}, S., \& {Zahn}, J.-P. 1992, \aap, 255, 191

\bibitem[{{Chayer} {et~al.}(1995){Chayer}, {Fontaine}, \&
  {Wesemael}}]{chayer95}
{Chayer}, P., {Fontaine}, G., \& {Wesemael}, F. 1995, \apjs, 99, 189

\bibitem[{{Deal} {et~al.}(2013){Deal}, {Vauclair}, \& {Vauclair}}]{deal13}
{Deal}, M., {Vauclair}, S., \& {Vauclair}, G. 2013, in Astronomical Society of
  the Pacific Conference Series, ed. J.~{Krzesi{\'n}ski}, G.~{Stachowski},
  P.~{Moskalik}, \& K.~{Bajan}, Vol. 469, 435

\bibitem[{{Desharnais} {et~al.}(2008){Desharnais}, {Wesemael}, {Chayer},
  {Kruk}, \& {Saffer}}]{desharnais08}
{Desharnais}, S., {Wesemael}, F., {Chayer}, P., {Kruk}, J.~W., \& {Saffer},
  R.~A. 2008, \apj, 672, 540

\bibitem[{{Dolez} \& {Vauclair}(1981)}]{dolez81}
{Dolez}, N. \& {Vauclair}, G. 1981, \aap, 102, 375

\bibitem[{{Dufour} {et~al.}(2012){Dufour}, {Kilic}, {Fontaine}, {Bergeron},
  {Melis}, \& {Bochanski}}]{dufour12}
{Dufour}, P., {Kilic}, M., {Fontaine}, G., {et~al.} 2012, \apj, 749, 6

\bibitem[{{Dupuis} {et~al.}(1992){Dupuis}, {Fontaine}, {Pelletier}, \&
  {Wesemael}}]{dupuis92I}
{Dupuis}, J., {Fontaine}, G., {Pelletier}, C., \& {Wesemael}, F. 1992, \apjs,
  82, 505

\bibitem[{{Dupuis} {et~al.}(1993{\natexlab{a}}){Dupuis}, {Fontaine},
  {Pelletier}, \& {Wesemael}}]{dupuis93II}
{Dupuis}, J., {Fontaine}, G., {Pelletier}, C., \& {Wesemael}, F.
  1993{\natexlab{a}}, \apjs, 84, 73

\bibitem[{{Dupuis} {et~al.}(1993{\natexlab{b}}){Dupuis}, {Fontaine}, \&
  {Wesemael}}]{dupuis93III}
{Dupuis}, J., {Fontaine}, G., \& {Wesemael}, F. 1993{\natexlab{b}}, \apjs, 87,
  345

\bibitem[{{Farihi}(2011)}]{farihi11}
{Farihi}, J. 2011, in American Institute of Physics Conference Series, Vol.
  1331, American Institute of Physics Conference Series, ed. S.~{Schuh},
  H.~{Drechsel}, \& U.~{Heber}, 193--210

\bibitem[{{Farihi} {et~al.}(2012){Farihi}, {G{\"a}nsicke}, {Wyatt}, {Girven},
  {Pringle}, \& {King}}]{farihi12}
{Farihi}, J., {G{\"a}nsicke}, B.~T., {Wyatt}, M.~C., {et~al.} 2012, \mnras,
  424, 464

\bibitem[{{Garaud}(2011)}]{garaud11}
{Garaud}, P. 2011, \apjl, 728, L30

\bibitem[{{Gill}(1982)}]{gill82}
{Gill}, A.~E. 1982, {Atmosphere-Ocean Dynamics, in International Geophysics
  Series, Vol. 30, ed. Academic Press}

\bibitem[{{Girven} {et~al.}(2012){Girven}, {Brinkworth}, {Farihi},
  {G{\"a}nsicke}, {Hoard}, {Marsh}, \& {Koester}}]{girven12}
{Girven}, J., {Brinkworth}, C.~S., {Farihi}, J., {et~al.} 2012, \apj, 749, 154

\bibitem[{{Huppert} \& {Moore}(1976)}]{huppert76}
{Huppert}, H.~E. \& {Moore}, D.~R. 1976, Journal of Fluid Mechanics, 78, 821

\bibitem[{{Itoh}(1992)}]{itoh92}
{Itoh}, N. 1992, \rmxaa, 23, 231

\bibitem[{{Jura}(2003)}]{jura03}
{Jura}, M. 2003, \apjl, 584, L91

\bibitem[{{Klein} {et~al.}(2010){Klein}, {Jura}, {Koester}, {Zuckerman}, \&
  {Melis}}]{klein10a}
{Klein}, B., {Jura}, M., {Koester}, D., {Zuckerman}, B., \& {Melis}, C. 2010,
  \apj, 709, 950

\bibitem[{{Koester}(2009)}]{koester09}
{Koester}, D. 2009, \aap, 498, 517

\bibitem[{{Koester} {et~al.}(2005){Koester}, {Rollenhagen}, {Napiwotzki},
  {Voss}, {Christlieb}, {Homeier}, \& {Reimers}}]{koester05}
{Koester}, D., {Rollenhagen}, K., {Napiwotzki}, R., {et~al.} 2005, \aap, 432,
  1025

\bibitem[{{Koester} \& {Wilken}(2006)}]{koester06}
{Koester}, D. \& {Wilken}, D. 2006, \aap, 453, 1051

\bibitem[{{Melis} {et~al.}(2011){Melis}, {Farihi}, {Dufour}, {Zuckerman},
  {Burgasser}, {Bergeron}, {Bochanski}, \& {Simcoe}}]{melis11}
{Melis}, C., {Farihi}, J., {Dufour}, P., {et~al.} 2011, \apj, 732, 90

\bibitem[{{Th{\'e}ado} \& {Vauclair}(2012)}]{theado12bis}
{Th{\'e}ado}, S. \& {Vauclair}, S. 2012, \apj, 744, 123

\bibitem[{{Traxler} {et~al.}(2011){Traxler}, {Garaud}, \& {Stellmach}}]{TGS11}
{Traxler}, A., {Garaud}, P., \& {Stellmach}, S. 2011, \apjl, 728, L29

\bibitem[{{Vauclair} {et~al.}(1979){Vauclair}, {Vauclair}, \&
  {Greenstein}}]{vauclair79}
{Vauclair}, G., {Vauclair}, S., \& {Greenstein}, J.~L. 1979, \aap, 80, 79

\bibitem[{{Vauclair}(2004)}]{vauclair04}
{Vauclair}, S. 2004, \apj, 605, 874

\bibitem[{{Vauclair} \& {Th{\'e}ado}(2012)}]{vauclair12}
{Vauclair}, S. \& {Th{\'e}ado}, S. 2012, \apj, 753, 49

\bibitem[{{Vennes} {et~al.}(2010){Vennes}, {Kawka}, \& {N{\'e}meth}}]{venne10}
{Vennes}, S., {Kawka}, A., \& {N{\'e}meth}, P. 2010, \mnras, 404, L40

\bibitem[{{Xu} \& {Jura}(2012)}]{xu12}
{Xu}, S. \& {Jura}, M. 2012, \apj, 745, 88

\bibitem[{{Zuckerman} {et~al.}(2011){Zuckerman}, {Koester}, {Dufour}, {Melis},
  {Klein}, \& {Jura}}]{zuckerman11}
{Zuckerman}, B., {Koester}, D., {Dufour}, P., {et~al.} 2011, \apj, 739, 101

\bibitem[{{Zuckerman} {et~al.}(2007){Zuckerman}, {Koester}, {Melis}, {Hansen},
  \& {Jura}}]{zuckerman07}
{Zuckerman}, B., {Koester}, D., {Melis}, C., {Hansen}, B.~M., \& {Jura}, M.
  2007, \apj, 671, 872

\bibitem[{{Zuckerman} {et~al.}(2003){Zuckerman}, {Koester}, {Reid}, \&
  {H{\"u}nsch}}]{zuckerman03}
{Zuckerman}, B., {Koester}, D., {Reid}, I.~N., \& {H{\"u}nsch}, M. 2003, \apj,
  596, 477

\bibitem[{{Zuckerman} {et~al.}(2010){Zuckerman}, {Melis}, {Klein}, {Koester},
  \& {Jura}}]{zuckerman10}
{Zuckerman}, B., {Melis}, C., {Klein}, B., {Koester}, D., \& {Jura}, M. 2010,
  \apj, 722, 725

\end{thebibliography}

\end{document}